\documentclass[aps,pra,oneclumn,groupedaddress,nofootinbib,longbibliography]{revtex4-2}
\pdfoutput=1

\usepackage{graphicx}
\usepackage{amsmath,amssymb,amsfonts}
\usepackage[colorlinks,citecolor=blue,linkcolor=blue,urlcolor=blue, breaklinks=true]{hyperref}

\usepackage{physics} 
\usepackage[percent]{overpic} 

\usepackage{xspace} 

\usepackage[english]{babel}

\usepackage{tikzit}

\tikzstyle{trapezium}=[fill=white, draw=black, shape=trapezium, trapezium stretches=true, trapezium right angle=-75, trapezium left angle=0, minimum width=1cm, minimum height=0.5cm, line width=0.025cm]
\tikzstyle{bigtrapezium}=[fill=white, draw=black, shape=trapezium, trapezium stretches=true, trapezium right angle=-75, trapezium left angle=0, minimum width=1.6cm, line width=0.025cm]
\tikzstyle{BIGtrapezium}=[fill=white, draw=black, shape=trapezium, trapezium stretches=true, trapezium right angle=-75, trapezium left angle=0, minimum width=1.8cm, line width=0.025cm]
\tikzstyle{daggerbigtrapezium}=[fill=white, draw=black, shape=trapezium, trapezium stretches=true, trapezium right angle=75, trapezium left angle=0, minimum width=1.6cm, line width=0.025cm]
\tikzstyle{state}=[fill=white, draw=black, shape=isosceles triangle, isosceles triangle stretches=true, inner sep=0.2, shape border rotate=-90, isosceles triangle apex angle=75, minimum width=0.7cm, line width=0.025cm, shape border uses incircle]
\tikzstyle{effect}=[fill=white, draw=black, shape=isosceles triangle, isosceles triangle stretches=false, inner sep=0.2, shape border rotate=90, isosceles triangle apex angle=75, minimum width=0.7cm, line width=0.025cm]
\tikzstyle{discard}=[shape=tlground, fill=white, draw=black, rotate=180, scale=1.2]
\tikzstyle{bistate}=[fill=white, draw=black, shape=isosceles triangle, isosceles triangle stretches=true, inner sep=0, shape border rotate=-90, isosceles triangle apex angle=95, minimum width=1cm, line width=0.025cm]
\tikzstyle{identity}=[shape=tlground, fill=white, draw=black, scale=1.2]
\tikzstyle{SAdiscard}=[shape=ground, fill=white, draw=black, rotate=180]
\tikzstyle{SAidentity}=[shape=ground, fill=white, draw=black]
\tikzstyle{square}=[fill=white, draw=black, shape=rectangle, minimum width=0.7cm, minimum height=0.7cm, line width=0.025cm, inner sep=0]
\tikzstyle{rectangle}=[fill=white, draw=black, shape=rectangle, minimum width=1cm, minimum height=0.6cm, line width=0.025cm, inner sep=0]
\tikzstyle{longrectangle}=[fill=white, draw=black, shape=rectangle, minimum width=1.5cm, minimum height=0.7cm, line width=0.025cm]
\tikzstyle{looongrectangle}=[fill=white, draw=black, shape=rectangle, minimum width=3cm, minimum height=0.7cm, line width=0.025cm]

\tikzstyle{black}=[-, draw=black, line width=0.025cm]
\tikzstyle{da}=[-, dashed, gray]
\tikzstyle{arrow}=[->, draw=black, line width=0.025cm]
\tikzstyle{gray}=[-, draw=gray]
\tikzstyle{doublearrow}=[<->, line width=0.025cm, draw=black]

\usepackage[]{circuitikzgit}

\newcommand{\cA}{\ensuremath{{\mathcal A}}\xspace}
\newcommand{\cB}{\ensuremath{{\mathcal B}}\xspace}

\begin{document}

\title{On the consistency of relative facts}

\author{Eric G. Cavalcanti}
\email{e.cavalcanti@griffith.edu.au (corresponding author)}
\affiliation{Centre for Quantum Dynamics, Griffith University, Yugambeh Country, Gold Coast, QLD 4222, Australia}
\author{Andrea {Di Biagio}}
\email{andrea.dibiagio@oeaw.ac.at}
\affiliation{Institute for Quantum Optics and Quantum Information (IQOQI) Vienna,
Austrian Academy of Sciences, Boltzmanngasse 3, A-1090 Vienna, Austria}
\affiliation{Basic Research Community for Physics e.V., Mariannenstraße 89, Leipzig, Germany}
\author{Carlo Rovelli}
\email{crovelli@uwo.ca}
\affiliation{Aix-Marseille University, Universit\'e de Toulon, CPT-CNRS, Marseille, France,}
\affiliation{Department of Philosophy and Rotman Institute of Philosophy, Western University, London ON, Canada,}
\affiliation{Perimeter Institute, 31 Caroline Street N, Waterloo ON, Canada}

\date{\small\today}

\begin{abstract} \noindent  
Lawrence \textit{et al.}~have presented an argument purporting to show that ``relative facts do not exist'' and, consequently,   ``Relational Quantum Mechanics is incompatible with quantum mechanics''. The argument is based on a GHZ-like contradiction between constraints satisfied by measurement outcomes in an extended Wigner's friend scenario. Here we present a strengthened version of the argument, and show why, contrary to the claim by Lawrence {\em et al.}, these arguments do not contradict the consistency of a theory of relative facts. Rather, considering this argument helps clarify how one should \emph{not} think about a theory of relative facts, like RQM.
\end{abstract}

\maketitle 

\raggedbottom

In \cite{lawrence2022relative}, Lawrence, Markiewicz, and {\.Z}ukowski present an argument meant to show that ``relative facts do not exist" and ``Relational Quantum Mechanics (RQM)  \cite{rovelli1996relational,rovelli2018space,dibiagio2021stable} is incompatible with Quantum Mechanics".  See also \cite{Drezet,LMZ2}. Here we show why their conclusion is not warranted. We also present a strengthened version of the argument and argue that,  although these arguments do not establish the inconsistency of relative facts, they nonetheless help clarify how one should \emph{not} think about a theory of relative facts, like RQM.

The authors consider an extended Wigner's friend version of a GHZ-type scenario. A system  $S$ formed by three qubits $(S_1,S_2,S_3)$ is prepared in a GHZ state  \cite{greenberger1990bell}: 
\begin{equation}
  \scalebox{1}{\begin{tikzpicture}
	\begin{pgfonlayer}{nodelayer}
		\node [style=none] (1) at (-2, 2) {};
		\node [style=none] (2) at (0, 2) {};
		\node [style=none] (3) at (2, 2) {};
		\node [style=none] (4) at (-1, -1.5) {};
		\node [style=none] (5) at (1, -1.5) {};
		\node [style=bistate, text width=0.8cm, align=center, minimum height=0.8cm] (6) at (0, -1.5) {\scalebox{0.75}{\raisebox{-0.5cm}{\makebox[0cm]{$\psi^{(S)}_\mathrm{GHZ}$}}}};
		\node [style=none] (7) at (-2.25, 1) {$S_1$};
		\node [style=none] (8) at (-0.5, 1) {$S_2$};
		\node [style=none] (9) at (1.25, 1) {$S_3$};
	\end{pgfonlayer}
	\begin{pgfonlayer}{edgelayer}
		\draw [style=black, in=90, out=-90] (1.center) to (4.center);
		\draw [style=black, in=90, out=-90] (3.center) to (5.center);
		\draw [style=black] (2.center) to (6);
	\end{pgfonlayer}
\end{tikzpicture}}
\end{equation}
A triple of systems $(A_1,A_2,A_3)$, considered as observers, respectively measure a fixed observable for each of the three qubits and obtains outcomes $(\mathcal {\cA}_1, {\cA}_2, {\cA}_3)$, where $i=1,2,3$. Subsequently, a second triple of observer systems $(B_1,B_2,B_3)$ respectively measures one observable for each pair of systems $(S_i, A_i)$ and obtains outcomes $({\cB}_1, {\cB}_2, {\cB}_3)$. 

Let us emphasise that, although the notation of \cite{lawrence2022relative} does not make this clear, in RQM these outcomes have a value relative to each observer being considered, but not necessarily relative to every observer. In other words, they are relative facts.

The observables are chosen in \cite{lawrence2022relative} to parallel the proof of the GHZ theorem \cite{greenberger1990bell} against the existence of local hidden variables. The authors of \cite{lawrence2022relative} claim that the resulting set of six quantities $\{\cA_1,\cA_2,\cA_3,\cB_1,\cB_2,\cB_3\}$ must satisfy the four incompatible GHZ constraints
\begin{eqnarray}
 (\mathrm{i}) &:& \ \ \ \  {\cB}_1{\cB}_2{\cB}_3=1, \nonumber \\
 (\mathrm{ii}) &:& \ \ \ \  {\cB}_1{\cA}_2{\cA}_3 = -1,  \nonumber \\ 
 (\mathrm{iii}) &:& \ \ \ \  {\cA}_1{\cB}_2{\cA}_3 = -1,  \nonumber \\
 (\mathrm{iv}) &:& \ \ \ \  {\cA}_1{\cA}_2{\cB}_3 = -1. \nonumber
 \end{eqnarray}
They conclude that this is an argument against the existence of the relative facts that RQM takes as its main ingredient.

Let us analyse this argument in detail. The authors of \cite{lawrence2022relative} argue that each of the above measurements can be described as a unitary interaction between the system being measured and the corresponding observer system (what they call an ``RQM-measurement''). This is in agreement with RQM.  However, in RQM, quantum states are only interpreted as \emph{relative} states in the sense of \cite{Everett:1957hd}, and therefore any such unitary evolution is relative to a specific observer. In \cite{lawrence2022relative}, the observer from whose perspective the unitary description is given is not made explicit. However, in RQM a state of a system is always a state relative to another system. Here we consider for concreteness the unitary description to be relative to an observer $W$ (``Wigner'') external to all of the systems considered above.

Using a notation only slightly different from that of \cite{lawrence2022relative}, the measurement of the Pauli $Y$ observable of $S_m$ by $A_m$ can be described as unitary $\hat{U}_{{SA}_m}$ such that, when $S_m$ is prepared in a $Y$-eigenstate $|l^y\rangle_{S_m}$ ($l\in\{\pm 1\}$) and $A_m$ is initially in a ``ready'' state $|R\rangle_{A_m}$, we have 
\begin{equation}
\hat{U}_{{SA}_m} \left(|l^y\rangle_{S_m} |R\rangle_{A_m} \right)=|l^y\rangle_{S_m} |l^y\rangle_{A_m}\,.
\end{equation}

 Ref.~\cite{lawrence2022relative} considers then an entangling measurement by $B_m$ on the joint system $S_m\otimes A_m$. 
 For simplicity of exposition, here (following \cite{bong2020strong}) we consider instead a measurement that consists of first applying the inverse unitary $\hat{U}_{{SA}_m}^\dagger$, and then $B_m$ proceeding to measure the system $S_m$ directly on the Pauli $X$ basis. This procedure leads to the same statistics.  
 
 Using this, Wigner describes the measurement by $B_m$ as a unitary interaction $\hat{U}_{{SAB}_m} = \hat{U}_{{SB}_m}\hat{U}_{{SA}_m}^\dagger$, where
\begin{equation}
\hat{U}_{{SB}_m} \left(|l^x\rangle_{S_m} |R\rangle_{B_m} \right)=|l^x\rangle_{S_m} |l^x\rangle_{B_m}\,.
\end{equation}
The sequence of all measurements then can be represented from Wigner's perspective as follows,  
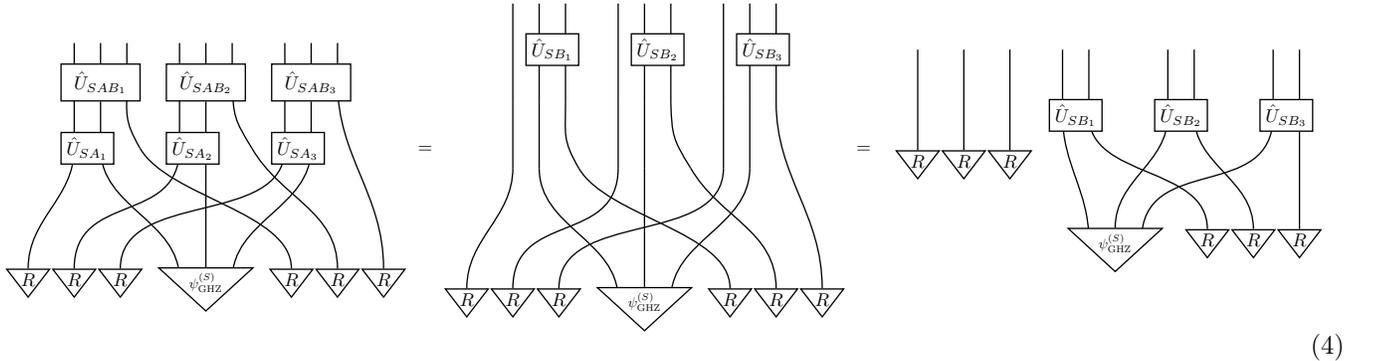
\begin{widetext}
\begin{equation}
\!\!\!\scalebox{0.7}{
    \begin{tikzpicture}
	\begin{pgfonlayer}{nodelayer}
		\node [style=none] (0) at (-4, 0) {};
		\node [style=none] (1) at (0, 0) {};
		\node [style=none] (2) at (4, 0) {};
		\node [style=none] (3) at (-1, -5) {};
		\node [style=none] (4) at (1, -5) {};
		\node [style=bistate, text width=0.8cm, align=center, minimum height=0.8cm] (5) at (0, -5) {\scalebox{0.75}{\raisebox{-0.5cm}{\makebox[0cm]{$\psi^{(S)}_\mathrm{GHZ}$}}}};
		\node [style=rectangle, minimum width=1cm] (6) at (-4.5, 0) {\makebox[0cm]{$\hat U_{SA_1}$}};
		\node [style=rectangle, minimum width=1cm] (7) at (-0.5, 0) {\makebox[0cm]{$\hat U_{SA_2}$}};
		\node [style=rectangle, minimum width=1cm] (8) at (3.5, 0) {\makebox[0cm]{$\hat U_{SA_3}$}};
		\node [style=state] (9) at (-5, -5) {$R$};
		\node [style=state] (10) at (-6.75, -5) {$R$};
		\node [style=state] (11) at (-3.25, -5) {$R$};
		\node [style=none] (12) at (-5, 0) {};
		\node [style=none] (13) at (-1, 0) {};
		\node [style=none] (14) at (3, 0) {};
		\node [style=none] (15) at (-5, 2.5) {};
		\node [style=none] (16) at (-1, 2.5) {};
		\node [style=none] (17) at (3, 2.5) {};
		\node [style=state] (18) at (5, -5) {$R$};
		\node [style=state] (19) at (3.25, -5) {$R$};
		\node [style=state] (20) at (6.75, -5) {$R$};
		\node [style=none] (21) at (5, 2.5) {};
		\node [style=longrectangle] (22) at (4, 2.5) {\makebox[0cm]{$\hat U_{SAB_3}$}};
		\node [style=longrectangle] (23) at (-4, 2.5) {\makebox[0cm]{$\hat U_{SAB_1}$}};
		\node [style=longrectangle] (24) at (0, 2.5) {\makebox[0cm]{$\hat U_{SAB_2}$}};
		\node [style=none] (25) at (-3, 2.5) {};
		\node [style=none] (26) at (1, 2.5) {};
		\node [style=none] (27) at (-2.5, 0) {};
		\node [style=none] (28) at (1.75, 0) {};
		\node [style=none] (29) at (-5, 4) {};
		\node [style=none] (30) at (-4, 4) {};
		\node [style=none] (31) at (-3, 4) {};
		\node [style=none] (32) at (-1, 4) {};
		\node [style=none] (33) at (0, 4) {};
		\node [style=none] (34) at (1, 4) {};
		\node [style=none] (35) at (3, 4) {};
		\node [style=none] (36) at (4, 4) {};
		\node [style=none] (37) at (5, 4) {};
	\end{pgfonlayer}
	\begin{pgfonlayer}{edgelayer}
		\draw [style=black, in=90, out=-90] (0.center) to (3.center);
		\draw [style=black, in=90, out=-90] (2.center) to (4.center);
		\draw [style=black, in=90, out=-90] (1.center) to (5);
		\draw [style=black, in=-90, out=90] (10) to (12.center);
		\draw [style=black, in=-90, out=90, looseness=1.25] (9) to (13.center);
		\draw [style=black, in=-90, out=90] (11) to (14.center);
		\draw [style=black] (12.center) to (15.center);
		\draw [style=black] (13.center) to (16.center);
		\draw [style=black] (14.center) to (17.center);
		\draw [style=black, in=90, out=-90] (21.center) to (20);
		\draw [style=black] (1.center) to (24);
		\draw [style=black] (0.center) to (23);
		\draw [style=black] (2.center) to (22);
		\draw [style=black, in=-90, out=120, looseness=1.25] (27.center) to (25.center);
		\draw [style=black, in=90, out=-60, looseness=0.75] (27.center) to (19);
		\draw [style=black, in=90, out=-60] (28.center) to (18);
		\draw [style=black, bend left=15, looseness=1.25] (28.center) to (26.center);
		\draw [style=black] (29.center) to (15.center);
		\draw [style=black] (30.center) to (23);
		\draw [style=black] (31.center) to (25.center);
		\draw [style=black] (32.center) to (16.center);
		\draw [style=black] (33.center) to (24);
		\draw [style=black] (34.center) to (26.center);
		\draw [style=black] (35.center) to (17.center);
		\draw [style=black] (36.center) to (22);
		\draw [style=black] (37.center) to (21.center);
	\end{pgfonlayer}
\end{tikzpicture}
    = \begin{tikzpicture}
	\begin{pgfonlayer}{nodelayer}
		\node [style=none] (0) at (-4, -0.75) {};
		\node [style=none] (1) at (0, -0.75) {};
		\node [style=none] (2) at (4, -0.75) {};
		\node [style=none] (3) at (-1, -5.75) {};
		\node [style=none] (4) at (1, -5.75) {};
		\node [style=bistate, text width=0.8cm, align=center, minimum height=0.8cm] (5) at (0, -5.75) {\scalebox{0.75}{\raisebox{-0.5cm}{\makebox[0cm]{$\psi^{(S)}_\mathrm{GHZ}$}}}};
		\node [style=state] (9) at (-5, -5.75) {$R$};
		\node [style=state] (10) at (-6.75, -5.75) {$R$};
		\node [style=state] (11) at (-3.25, -5.75) {$R$};
		\node [style=none] (12) at (-5, -0.75) {};
		\node [style=none] (13) at (-1, -0.75) {};
		\node [style=none] (14) at (3, -0.75) {};
		\node [style=none] (15) at (-5, 1.75) {};
		\node [style=none] (16) at (-1, 1.75) {};
		\node [style=none] (17) at (3, 1.75) {};
		\node [style=state] (18) at (5, -5.75) {$R$};
		\node [style=state] (19) at (3.25, -5.75) {$R$};
		\node [style=state] (20) at (6.75, -5.75) {$R$};
		\node [style=none] (21) at (5, 1.75) {};
		\node [style=none] (22) at (4, 1.75) {};
		\node [style=none] (23) at (-4, 1.75) {};
		\node [style=none] (24) at (0, 1.75) {};
		\node [style=none] (25) at (-3, 1.75) {};
		\node [style=none] (26) at (1, 1.75) {};
		\node [style=none] (27) at (-2.5, -0.75) {};
		\node [style=none] (28) at (1.75, -0.75) {};
		\node [style=none] (29) at (-5, 5.5) {};
		\node [style=none] (30) at (-4, 5.5) {};
		\node [style=none] (31) at (-3, 5.5) {};
		\node [style=none] (32) at (-1, 5.5) {};
		\node [style=none] (33) at (0, 5.5) {};
		\node [style=none] (34) at (1, 5.5) {};
		\node [style=none] (35) at (3, 5.5) {};
		\node [style=none] (36) at (4, 5.5) {};
		\node [style=none] (37) at (5, 5.5) {};
		\node [style=rectangle, minimum width=1cm] (41) at (-3.5, 3.75) {\makebox[0cm]{$\hat U_{SB_1}$}};
		\node [style=rectangle, minimum width=1cm] (42) at (0.5, 3.75) {\makebox[0cm]{$\hat U_{SB_2}$}};
		\node [style=rectangle, minimum width=1cm] (43) at (4.5, 3.75) {\makebox[0cm]{$\hat U_{SB_3}$}};
	\end{pgfonlayer}
	\begin{pgfonlayer}{edgelayer}
		\draw [style=black, in=90, out=-90] (0.center) to (3.center);
		\draw [style=black, in=90, out=-90] (2.center) to (4.center);
		\draw [style=black, in=90, out=-90] (1.center) to (5);
		\draw [style=black, in=-90, out=90] (10) to (12.center);
		\draw [style=black, in=-90, out=90, looseness=1.25] (9) to (13.center);
		\draw [style=black, in=-90, out=90] (11) to (14.center);
		\draw [style=black] (12.center) to (15.center);
		\draw [style=black] (13.center) to (16.center);
		\draw [style=black] (14.center) to (17.center);
		\draw [style=black, in=90, out=-90] (21.center) to (20);
		\draw [style=black] (1.center) to (24.center);
		\draw [style=black] (0.center) to (23.center);
		\draw [style=black] (2.center) to (22.center);
		\draw [style=black, in=-90, out=120, looseness=1.25] (27.center) to (25.center);
		\draw [style=black, in=90, out=-60, looseness=0.75] (27.center) to (19);
		\draw [style=black, in=90, out=-60] (28.center) to (18);
		\draw [style=black, bend left=15, looseness=1.25] (28.center) to (26.center);
		\draw [style=black] (29.center) to (15.center);
		\draw [style=black] (30.center) to (23.center);
		\draw [style=black] (31.center) to (25.center);
		\draw [style=black] (32.center) to (16.center);
		\draw [style=black] (33.center) to (24.center);
		\draw [style=black] (34.center) to (26.center);
		\draw [style=black] (35.center) to (17.center);
		\draw [style=black] (36.center) to (22.center);
		\draw [style=black] (37.center) to (21.center);
	\end{pgfonlayer}
\end{tikzpicture} 
    =~~ \begin{tikzpicture}
	\begin{pgfonlayer}{nodelayer}
		\node [style=none] (0) at (-4.5, 1.25) {};
		\node [style=none] (1) at (-0.5, 1.25) {};
		\node [style=none] (2) at (3.5, 1.25) {};
		\node [style=none] (3) at (-3.5, -3.5) {};
		\node [style=none] (4) at (-1.5, -3.5) {};
		\node [style=bistate, text width=0.8cm, align=center, minimum height=0.8cm] (5) at (-2.5, -3.5) {\scalebox{0.75}{\raisebox{-0.5cm}{\makebox[0cm]{$\psi^{(S)}_\mathrm{GHZ}$}}}};
		\node [style=state] (9) at (-8.25, -0.5) {$R$};
		\node [style=state] (10) at (-10, -0.5) {$R$};
		\node [style=state] (11) at (-6.5, -0.5) {$R$};
		\node [style=none] (12) at (-10, 3.75) {};
		\node [style=none] (13) at (-8.25, 3.75) {};
		\node [style=none] (14) at (-6.5, 3.75) {};
		\node [style=state] (18) at (2.75, -3.5) {$R$};
		\node [style=state] (19) at (1, -3.5) {$R$};
		\node [style=state] (20) at (4.5, -3.5) {$R$};
		\node [style=none] (21) at (4.5, 1.25) {};
		\node [style=none] (27) at (-3.5, 1.25) {};
		\node [style=none] (28) at (0.5, 1.25) {};
		\node [style=rectangle, minimum width=1cm] (41) at (-4, 1.25) {\makebox[0cm]{$\hat U_{SB_1}$}};
		\node [style=rectangle, minimum width=1cm] (42) at (0, 1.25) {\makebox[0cm]{$\hat U_{SB_2}$}};
		\node [style=rectangle, minimum width=1cm] (43) at (4, 1.25) {\makebox[0cm]{$\hat U_{SB_3}$}};
		\node [style=none] (44) at (-4.5, 3.75) {};
		\node [style=none] (45) at (-3.5, 3.75) {};
		\node [style=none] (46) at (-0.5, 3.75) {};
		\node [style=none] (47) at (0.5, 3.75) {};
		\node [style=none] (48) at (3.5, 3.75) {};
		\node [style=none] (49) at (4.5, 3.75) {};
	\end{pgfonlayer}
	\begin{pgfonlayer}{edgelayer}
		\draw [style=black, in=90, out=-90] (0.center) to (3.center);
		\draw [style=black, in=90, out=-90, looseness=1.25] (2.center) to (4.center);
		\draw [style=black, in=90, out=-90] (1.center) to (5);
		\draw [style=black, in=-90, out=90] (10) to (12.center);
		\draw [style=black, in=-90, out=90, looseness=1.25] (9) to (13.center);
		\draw [style=black, in=-90, out=90] (11) to (14.center);
		\draw [style=black, in=90, out=-90] (21.center) to (20);
		\draw [style=black, in=90, out=-90] (27.center) to (19);
		\draw [style=black, in=90, out=-90] (28.center) to (18);
		\draw [style=black] (44.center) to (0.center);
		\draw [style=black] (45.center) to (27.center);
		\draw [style=black] (46.center) to (1.center);
		\draw [style=black] (47.center) to (28.center);
		\draw [style=black] (48.center) to (2.center);
		\draw [style=black] (49.center) to (21.center);
	\end{pgfonlayer}
\end{tikzpicture}
    }
\end{equation}
\end{widetext}
where we have used
\begin{equation}
  \scalebox{0.8}{  \begin{tikzpicture}
	\begin{pgfonlayer}{nodelayer}
		\node [style=none] (0) at (0, -2.5) {};
		\node [style=rectangle, minimum width=1cm] (6) at (-0.5, -1) {\makebox[0cm]{$\hat U_{SA_m}$}};
		\node [style=none] (12) at (-1, -2.5) {};
		\node [style=none] (15) at (-1, 1.5) {};
		\node [style=longrectangle] (23) at (0, 1.5) {\makebox[0cm]{$\hat U_{SAB_m}$}};
		\node [style=none] (25) at (1, 1.5) {};
		\node [style=none] (27) at (1, -2.5) {};
		\node [style=none] (29) at (-1, 3) {};
		\node [style=none] (30) at (0, 3) {};
		\node [style=none] (31) at (1, 3) {};
	\end{pgfonlayer}
	\begin{pgfonlayer}{edgelayer}
		\draw [style=black] (12.center) to (15.center);
		\draw [style=black] (0.center) to (23);
		\draw [style=black] (27.center) to (25.center);
		\draw [style=black] (29.center) to (15.center);
		\draw [style=black] (30.center) to (23);
		\draw [style=black] (31.center) to (25.center);
	\end{pgfonlayer}
\end{tikzpicture} ~~~=~~~ \begin{tikzpicture}
	\begin{pgfonlayer}{nodelayer}
		\node [style=none] (0) at (0, -4) {};
		\node [style=rectangle, minimum width=1cm] (6) at (-0.5, -2.5) {\makebox[0cm]{$\hat U_{SA_m}$}};
		\node [style=none] (12) at (-1, -4) {};
		\node [style=none] (15) at (-1, 0) {};
		\node [style=none] (23) at (0, 0) {};
		\node [style=none] (25) at (1, 0) {};
		\node [style=none] (27) at (1, -4) {};
		\node [style=none] (29) at (-1, 3.75) {};
		\node [style=none] (30) at (0, 3.75) {};
		\node [style=none] (31) at (1, 3.75) {};
		\node [style=rectangle, minimum width=1cm] (38) at (-0.5, 0) {\makebox[0cm]{$\hat U^\dagger_{SA_m}$}};
		\node [style=rectangle, minimum width=1cm] (41) at (0.5, 2) {\makebox[0cm]{$\hat U_{SB_m}$}};
	\end{pgfonlayer}
	\begin{pgfonlayer}{edgelayer}
		\draw [style=black] (12.center) to (15.center);
		\draw [style=black] (0.center) to (23.center);
		\draw [style=black] (27.center) to (25.center);
		\draw [style=black] (29.center) to (15.center);
		\draw [style=black] (30.center) to (23.center);
		\draw [style=black] (31.center) to (25.center);
	\end{pgfonlayer}
\end{tikzpicture} ~~~=~~~ \begin{tikzpicture}
	\begin{pgfonlayer}{nodelayer}
		\node [style=none] (0) at (0, -2) {};
		\node [style=none] (12) at (-1, -2) {};
		\node [style=none] (15) at (-1, 2) {};
		\node [style=none] (23) at (0, 2) {};
		\node [style=none] (25) at (1, 2) {};
		\node [style=none] (27) at (1, -2) {};
		\node [style=rectangle, minimum width=1cm] (41) at (0.5, 0) {\makebox[0cm]{$\hat U_{SB_m}$}};
	\end{pgfonlayer}
	\begin{pgfonlayer}{edgelayer}
		\draw [style=black] (12.center) to (15.center);
		\draw [style=black] (0.center) to (23.center);
		\draw [style=black] (27.center) to (25.center);
	\end{pgfonlayer}
\end{tikzpicture}}.
\end{equation}

Now let us consider the constraints (i)-(iv) above, starting with (i). The authors of \cite{lawrence2022relative} describe the composite system $B=B_1 \otimes B_2\otimes B_3$ as a single observer ``$B$'', and similarly for $A=A_1\otimes A_2 \otimes A_3$. But this is not necessarily coherent with RQM.

Let us first consider the three systems $B_i$ as separate observers. The outcome $\cB_i$ has a value relative to observer $B_m$, but can we say that the product of these outcomes should satisfy (i)? In RQM, as quoted in \cite{lawrence2022relative}, ``it is meaningless to compare events relative to different systems, unless this is done relative to a (possibly third) system'' and ``comparisons can only be made by a (quantum–mechanical) interaction''. Thus, before the observers $B_m$ interact among themselves, or with a further observer, the constraint (i) has no meaning in RQM.

Let us then consider an interaction with Wigner, who measures each system $B_m$ on its ``pointer basis''---that is, the basis $|l^x\rangle_{B_m}$ above, obtaining outcome $\cB_m^W$. 
It is easy to show that the 
statistics for the product of these three measurements should correspond to the statistics for the product of three Pauli $X$ measurements on the initial GHZ state. 
We can represent this process diagramatically as follows
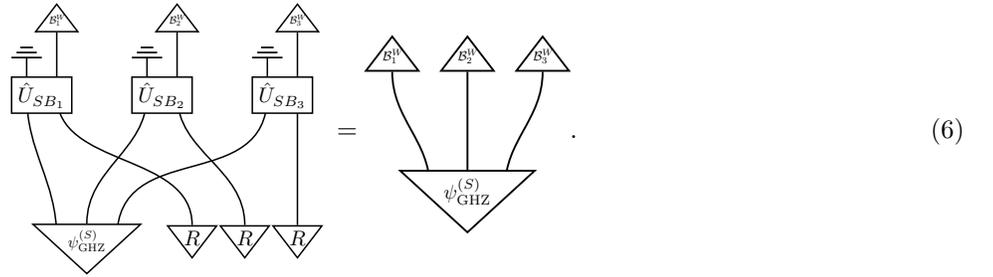
\begin{equation}
    \scalebox{0.8}{\begin{tikzpicture}
	\begin{pgfonlayer}{nodelayer}
		\node [style=none] (0) at (-4.5, 1.25) {};
		\node [style=none] (1) at (-0.5, 1.25) {};
		\node [style=none] (2) at (3.5, 1.25) {};
		\node [style=none] (3) at (-3.5, -3.5) {};
		\node [style=none] (4) at (-1.5, -3.5) {};
		\node [style=bistate, text width=0.8cm, align=center, minimum height=0.8cm] (5) at (-2.5, -3.5) {\scalebox{0.75}{\raisebox{-0.5cm}{\makebox[0cm]{$\psi^{(S)}_\mathrm{GHZ}$}}}};
		\node [style=state] (18) at (2.75, -3.5) {$R$};
		\node [style=state] (19) at (1, -3.5) {$R$};
		\node [style=state] (20) at (4.5, -3.5) {$R$};
		\node [style=none] (21) at (4.5, 1.25) {};
		\node [style=none] (27) at (-3.5, 1.25) {};
		\node [style=none] (28) at (0.5, 1.25) {};
		\node [style=rectangle, minimum width=1cm] (41) at (-4, 1.25) {\makebox[0cm]{$\hat U_{SB_1}$}};
		\node [style=rectangle, minimum width=1cm] (42) at (0, 1.25) {\makebox[0cm]{$\hat U_{SB_2}$}};
		\node [style=rectangle, minimum width=1cm] (43) at (4, 1.25) {\makebox[0cm]{$\hat U_{SB_3}$}};
		\node [style=discard] (44) at (-4.5, 2.5) {};
		\node [style=effect] (45) at (-3.5, 3.75) {\scalebox{0.50}{\makebox[0cm]{$\,\mathcal B_1^W$}}};
		\node [style=discard] (46) at (-0.5, 2.5) {};
		\node [style=effect] (47) at (0.5, 3.75) {\scalebox{0.50}{\makebox[0cm]{$\,\mathcal B_2^W$}}};
		\node [style=discard] (48) at (3.5, 2.5) {};
		\node [style=effect] (49) at (4.5, 3.75) {\scalebox{0.50}{\makebox[0cm]{$\,\mathcal B_3^W$}}};
	\end{pgfonlayer}
	\begin{pgfonlayer}{edgelayer}
		\draw [style=black, in=90, out=-90] (0.center) to (3.center);
		\draw [style=black, in=90, out=-90, looseness=1.25] (2.center) to (4.center);
		\draw [style=black, in=90, out=-90] (1.center) to (5);
		\draw [style=black, in=90, out=-90] (21.center) to (20);
		\draw [style=black, in=90, out=-90] (27.center) to (19);
		\draw [style=black, in=90, out=-90] (28.center) to (18);
		\draw [style=black] (44) to (0.center);
		\draw [style=black] (45) to (27.center);
		\draw [style=black] (46) to (1.center);
		\draw [style=black] (47) to (28.center);
		\draw [style=black] (48) to (2.center);
		\draw [style=black] (49) to (21.center);
	\end{pgfonlayer}
\end{tikzpicture}}=\begin{tikzpicture}
	\begin{pgfonlayer}{nodelayer}
		\node [style=effect] (1) at (-2, 2) {\scalebox{0.50}{\makebox[0cm]{$\,\mathcal B_1^W$}}};
		\node [style=effect] (2) at (0, 2) {\scalebox{0.5}{\makebox[0cm]{$\,\mathcal B_2^W$}}};
		\node [style=effect] (3) at (2, 2) {\scalebox{0.5}{\makebox[0cm]{$\,\mathcal B_3^W$}}};
		\node [style=none] (4) at (-1, -1.5) {};
		\node [style=none] (5) at (1, -1.5) {};
		\node [style=bistate, text width=0.8cm, align=center, minimum height=0.8cm] (6) at (0, -1.5) {\scalebox{0.75}{\raisebox{-0.5cm}{\makebox[0cm]{$\psi^{(S)}_\mathrm{GHZ}$}}}};
	\end{pgfonlayer}
	\begin{pgfonlayer}{edgelayer}
		\draw [style=black, in=90, out=-90] (1) to (4.center);
		\draw [style=black, in=90, out=-90] (3) to (5.center);
		\draw [style=black] (2) to (6);
	\end{pgfonlayer}
\end{tikzpicture}.
\end{equation}
This satisfies
\begin{eqnarray}
 (\mathrm{i'}) &:& \ \ \ \  {\cB}_1^W{\cB}_2^W{\cB}_3^W=1. \nonumber
 \end{eqnarray}
But this is not constraint (i), and in RQM we cannot infer constraint (i) from this constraint.

Similarly, constraint (ii) is not meaningful in RQM except relative to an observer that evaluates it. Let us then consider the situation where Wigner measures $B_1$ as above, but this time measures systems $A_2$ and $A_3$ on their pointer bases before $B_2$ and $B_3$ do their measurements, obtaining outcomes $(\cB_1^W,\cA_2^W, \cA_3^W)$:
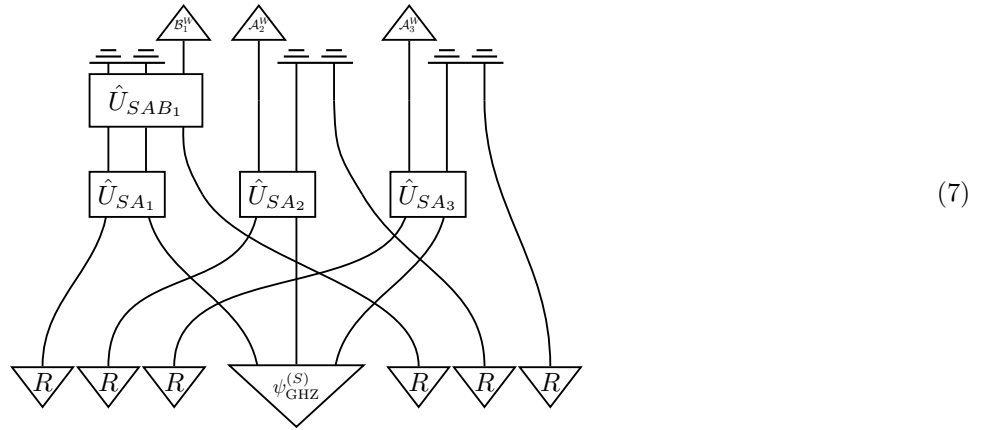
\begin{equation}
    \begin{tikzpicture}
	\begin{pgfonlayer}{nodelayer}
		\node [style=none] (0) at (-4, 0) {};
		\node [style=none] (1) at (0, 0) {};
		\node [style=none] (2) at (4, 0) {};
		\node [style=none] (3) at (-1, -5) {};
		\node [style=none] (4) at (1, -5) {};
		\node [style=bistate, text width=0.8cm, align=center, minimum height=0.8cm] (5) at (0, -5) {\scalebox{0.75}{\raisebox{-0.5cm}{\makebox[0cm]{$\psi^{(S)}_\mathrm{GHZ}$}}}};
		\node [style=rectangle, minimum width=1cm] (6) at (-4.5, 0) {\makebox[0cm]{$\hat U_{SA_1}$}};
		\node [style=rectangle, minimum width=1cm] (7) at (-0.5, 0) {\makebox[0cm]{$\hat U_{SA_2}$}};
		\node [style=rectangle, minimum width=1cm] (8) at (3.5, 0) {\makebox[0cm]{$\hat U_{SA_3}$}};
		\node [style=state] (9) at (-5, -5) {$R$};
		\node [style=state] (10) at (-6.75, -5) {$R$};
		\node [style=state] (11) at (-3.25, -5) {$R$};
		\node [style=none] (12) at (-5, 0) {};
		\node [style=none] (13) at (-1, 0) {};
		\node [style=none] (14) at (3, 0) {};
		\node [style=none] (15) at (-5, 2.5) {};
		\node [style=none] (16) at (-1, 2.5) {};
		\node [style=none] (17) at (3, 2.5) {};
		\node [style=state] (18) at (5, -5) {$R$};
		\node [style=state] (19) at (3.25, -5) {$R$};
		\node [style=state] (20) at (6.75, -5) {$R$};
		\node [style=none] (21) at (5, 2.5) {};
		\node [style=longrectangle] (23) at (-4, 2.5) {\makebox[0cm]{$\hat U_{SAB_1}$}};
		\node [style=none] (25) at (-3, 2.5) {};
		\node [style=none] (26) at (1, 2.5) {};
		\node [style=none] (27) at (-2.5, 0) {};
		\node [style=none] (28) at (1.75, 0) {};
		\node [style=discard] (29) at (-5, 3.5) {};
		\node [style=discard] (30) at (-4, 3.5) {};
		\node [style=effect] (31) at (-3, 4.5) {\scalebox{0.50}{\makebox[0cm]{$\,\mathcal B_1^W$}}};
		\node [style=effect] (32) at (-1, 4.5) {\scalebox{0.50}{\makebox[0cm]{$\,\mathcal A_2^W$}}};
		\node [style=discard] (33) at (0, 3.5) {};
		\node [style=discard] (34) at (1, 3.5) {};
		\node [style=effect] (35) at (3, 4.5) {\scalebox{0.50}{\makebox[0cm]{$\,\mathcal A_3^W$}}};
		\node [style=discard] (36) at (4, 3.5) {};
		\node [style=discard] (37) at (5, 3.5) {};
	\end{pgfonlayer}
	\begin{pgfonlayer}{edgelayer}
		\draw [style=black, in=90, out=-90] (0.center) to (3.center);
		\draw [style=black, in=90, out=-90] (2.center) to (4.center);
		\draw [style=black, in=90, out=-90] (1.center) to (5);
		\draw [style=black, in=-90, out=90] (10) to (12.center);
		\draw [style=black, in=-90, out=90, looseness=1.25] (9) to (13.center);
		\draw [style=black, in=-90, out=90] (11) to (14.center);
		\draw [style=black] (12.center) to (15.center);
		\draw [style=black] (13.center) to (16.center);
		\draw [style=black] (14.center) to (17.center);
		\draw [style=black, in=90, out=-90] (21.center) to (20);
		\draw [style=black] (0.center) to (23);
		\draw [style=black, in=-90, out=120, looseness=1.25] (27.center) to (25.center);
		\draw [style=black, in=90, out=-60, looseness=0.75] (27.center) to (19);
		\draw [style=black, in=90, out=-60] (28.center) to (18);
		\draw [style=black, bend left=15, looseness=1.25] (28.center) to (26.center);
		\draw [style=black] (29) to (15.center);
		\draw [style=black] (30) to (23);
		\draw [style=black] (31) to (25.center);
		\draw [style=black] (32) to (16.center);
		\draw [style=black] (34) to (26.center);
		\draw [style=black] (35) to (17.center);
		\draw [style=black] (37) to (21.center);
		\draw [style=black] (33) to (1.center);
		\draw [style=black] (36) to (2.center);
	\end{pgfonlayer}
\end{tikzpicture}
\end{equation}
The statistics for the product of these three measurements correspond to the statistics for the product of Pauli measurements $X_1Y_2Y_3$ on the initial GHZ state, thus satisfying
\begin{eqnarray}
 (\mathrm{ii'}) &:& \ \ \ \  {\cB}_1^W{\cA}_2^W{\cA}_3^W=-1. \nonumber
 \end{eqnarray}
But as before, this is not constraint (ii), and in RQM we cannot infer constraint (ii) from this constraint. A similar analysis holds for constraints (iii) and (iv). 

We therefore conclude that \emph{none} of the constraints (i)-(iv) hold a priori in RQM, contrary to the claim by Ref.~\cite{lawrence2022relative}. Only one of the four constraints (i')-(iv') can hold relative to Wigner, with the rest being meaningless. Each constraint corresponds to a different context, where Wigner makes a different triple of measurements. To be clear, Wigner can meaningfully predict, before choosing his measurements, that the following constraints hold as expectation values, if those measurements are performed by him:
\begin{eqnarray}
 (\mathrm{i'''}) &:& \ \ \ \  \langle{\hat{X}}_1{\hat{X}}_2{\hat{X}}_3\rangle_W=1, \nonumber \\
 (\mathrm{ii'''}) &:& \ \ \ \  \langle{\hat{X}}_1{\hat{Y}}_2{\hat{Y}}_3\rangle_W = -1,  \nonumber \\ 
 (\mathrm{iii'''}) &:& \ \ \ \  \langle{\hat{Y}}_1{\hat{X}}_2{\hat{Y}}_3\rangle_W  = -1,  \nonumber \\
 (\mathrm{iv'''}) &:& \ \ \ \  \langle{\hat{Y}}_1{\hat{Y}}_2{\hat{X}}_3\rangle_W  = -1. \nonumber
 \end{eqnarray}
But when we write $\cB_1^W$, we are referring to a measurement outcome actually obtained by Wigner in a particular run. Of course, this value only exists (relative to Wigner) in the runs where Wigner performs that measurement. Following Peres \cite{Peres1978}, ``unperformed measurements have no outcomes".

On the other hand, one may ask: isn't it the case that all the six quantities $\{\cA_1,\cA_2,\cA_3,\cB_1,\cB_2,\cB_3\}$ refer to \emph{performed} measurements? Don't they all have a value in each run of the experiment, then? And if so, shouldn't those values obey the constraints (i)-(iv)? 

This is a subtle point. The key is that although those measurements are all performed \emph{by some observer} in each run of the experiment, there is no observer relative to whom they all take co-existing values. One may invoke the ``cross-perspective link'' \cite{adlam2022information} to conclude that \emph{if} Wigner performs one of the six measurements above (say if he observes outcome $\cB_1^W$), \emph{then} he can conclude that $\cB_1^W=\cB_1$. If he observes the triple $\cB_1^W\cB_2^W\cB_3^W$, he should obtain values compatible with (i'), and therefore in that case he could conclude that constraint (i) holds for the values observed by $B_1$, $B_2$ and $B_3$. One cannot however simply \emph{define} an observer $B = B_1 \otimes B_2 \otimes B_3$ relative to which constraint (i) holds, if there is no interaction involving those three systems after their measurements take place. A similar argument can be made to conclude that \emph{if} Wigner performs the measurements corresponding to one of the other constraints (ii')-(iv'), \emph{then} the corresponding constraint (ii)-(iv) holds. But this does not allow us to infer that all four constraints must be a priori satisfied.
 
We close with a general philosophical consideration.  As repeatedly stated in the original papers, RQM does not necessarily require a commitment to a specific philosophy.  However, it does highlight the cost that quantum mechanics puts upon different metaphysical options. The scenario analysed here is a good example. Different philosophical attitudes can be considered, with respect to the metaphysical status of the list  $\{\cA_1,\cA_2,\cA_3,\cB_1,\cB_2,\cB_3\}$. One possibility is the choice of declaring it part of reality, even if no observer has simultaneous access to all of those values (see \cite{adlam2022information}). The ``cost" of this option is that reality, so defined, violates a number of features that we commonly expect it to respect~\cite{brukner2018nogo, bong2020strong, cavalcanti2021implicationsa, Haddara2022} -- an assignment of values to all of those quantities amounts to an assumption of ``Absoluteness of Observed Events'', implying the rejection of at least one of the other premises of various no-go theorems~\cite{bong2020strong, cavalcanti2021implicationsa, Haddara2022}. Alternatively, one may choose a more radical relationalism, and assume that only assertions relative to a physical system are to be taken as meaningful statements about reality.  In this case, the elements of the list are part of reality relative to each observer making those measurements, but the complete list is not part of reality, because there is no observer relative to which all of those observables take co-existing values.

\centerline{***}

\begin{acknowledgments}
EGC acknowledges support from grant number FQXi-RFP-CPW-2019 from the Foundational Questions Institute and Fetzer Franklin Fund, a donor advised fund of Silicon Valley Community Foundation (EGC), and an Australian Research Council (ARC) Future Fellowship FT180100317 (EGC). ADB and CR acknowledge support of the ID\# 61466 grant from the John Templeton Foundation, as part of the ``Quantum Information Structure of Spacetime (QISS)'' project (\hyperlink{http://www.qiss.fr}{qiss.fr}).  EGC acknowledges the traditional owners of the land at Griffith University on which this work was undertaken, the Yuggera and Yugambeh peoples.
\end{acknowledgments}

\bibliography{list.bib}

\begin{thebibliography}{14}%
\makeatletter
\providecommand \@ifxundefined [1]{%
 \@ifx{#1\undefined}
}%
\providecommand \@ifnum [1]{%
 \ifnum #1\expandafter \@firstoftwo
 \else \expandafter \@secondoftwo
 \fi
}%
\providecommand \@ifx [1]{%
 \ifx #1\expandafter \@firstoftwo
 \else \expandafter \@secondoftwo
 \fi
}%
\providecommand \natexlab [1]{#1}%
\providecommand \enquote  [1]{``#1''}%
\providecommand \bibnamefont  [1]{#1}%
\providecommand \bibfnamefont [1]{#1}%
\providecommand \citenamefont [1]{#1}%
\providecommand \href@noop [0]{\@secondoftwo}%
\providecommand \href [0]{\begingroup \@sanitize@url \@href}%
\providecommand \@href[1]{\@@startlink{#1}\@@href}%
\providecommand \@@href[1]{\endgroup#1\@@endlink}%
\providecommand \@sanitize@url [0]{\catcode `\\12\catcode `\$12\catcode
  `\&12\catcode `\#12\catcode `\^12\catcode `\_12\catcode `\%12\relax}%
\providecommand \@@startlink[1]{}%
\providecommand \@@endlink[0]{}%
\providecommand \url  [0]{\begingroup\@sanitize@url \@url }%
\providecommand \@url [1]{\endgroup\@href {#1}{\urlprefix }}%
\providecommand \urlprefix  [0]{URL }%
\providecommand \Eprint [0]{\href }%
\providecommand \doibase [0]{https://doi.org/}%
\providecommand \selectlanguage [0]{\@gobble}%
\providecommand \bibinfo  [0]{\@secondoftwo}%
\providecommand \bibfield  [0]{\@secondoftwo}%
\providecommand \translation [1]{[#1]}%
\providecommand \BibitemOpen [0]{}%
\providecommand \bibitemStop [0]{}%
\providecommand \bibitemNoStop [0]{.\EOS\space}%
\providecommand \EOS [0]{\spacefactor3000\relax}%
\providecommand \BibitemShut  [1]{\csname bibitem#1\endcsname}%
\let\auto@bib@innerbib\@empty
\bibitem [{\citenamefont {Lawrence}\ \emph
  {et~al.}(2022{\natexlab{a}})\citenamefont {Lawrence}, \citenamefont
  {Markiewicz},\ and\ \citenamefont {{\.Z}ukowski}}]{lawrence2022relative}%
  \BibitemOpen
  \bibfield  {author} {\bibinfo {author} {\bibfnamefont {J.}~\bibnamefont
  {Lawrence}}, \bibinfo {author} {\bibfnamefont {M.}~\bibnamefont
  {Markiewicz}},\ and\ \bibinfo {author} {\bibfnamefont {M.}~\bibnamefont
  {{\.Z}ukowski}},\ }\href {http://arxiv.org/abs/2208.11793} {\bibinfo {title}
  {Relative facts do not exist. {{Relational Quantum Mechanics}} is
  {{Incompatible}} with {{Quantum Mechanics}}}} (\bibinfo {year}
  {2022}{\natexlab{a}}),\ \Eprint {https://arxiv.org/abs/2208.11793}
  {arXiv:2208.11793} \BibitemShut {NoStop}%
\bibitem [{\citenamefont {Rovelli}(1996)}]{rovelli1996relational}%
  \BibitemOpen
  \bibfield  {author} {\bibinfo {author} {\bibfnamefont {C.}~\bibnamefont
  {Rovelli}},\ }\bibfield  {title} {\bibinfo {title} {Relational {{Quantum
  Mechanics}}},\ }\href {https://doi.org/10/bx4mzr} {\bibfield  {journal}
  {\bibinfo  {journal} {International Journal of Theoretical Physics}\ }\textbf
  {\bibinfo {volume} {35}},\ \bibinfo {pages} {1637} (\bibinfo {year}
  {1996})},\ \Eprint {https://arxiv.org/abs/quant-ph/9609002}
  {arXiv:quant-ph/9609002} \BibitemShut {NoStop}%
\bibitem [{\citenamefont {Rovelli}(2018)}]{rovelli2018space}%
  \BibitemOpen
  \bibfield  {author} {\bibinfo {author} {\bibfnamefont {C.}~\bibnamefont
  {Rovelli}},\ }\bibfield  {title} {\bibinfo {title} {``{{Space}} is blue and
  birds fly through it''},\ }\href {https://doi.org/10/gks4cc} {\bibfield
  {journal} {\bibinfo  {journal} {Philosophical Transactions of the Royal
  Society A: Mathematical, Physical and Engineering Sciences}\ }\textbf
  {\bibinfo {volume} {376}},\ \bibinfo {pages} {20170312} (\bibinfo {year}
  {2018})},\ \Eprint {https://arxiv.org/abs/1712.02894} {arXiv:1712.02894}
  \BibitemShut {NoStop}%
\bibitem [{\citenamefont {Di~Biagio}\ and\ \citenamefont
  {Rovelli}(2021)}]{dibiagio2021stable}%
  \BibitemOpen
  \bibfield  {author} {\bibinfo {author} {\bibfnamefont {A.}~\bibnamefont
  {Di~Biagio}}\ and\ \bibinfo {author} {\bibfnamefont {C.}~\bibnamefont
  {Rovelli}},\ }\bibfield  {title} {\bibinfo {title} {Stable {{Facts}},
  {{Relative Facts}}},\ }\href {https://doi.org/10/gm7w6w} {\bibfield
  {journal} {\bibinfo  {journal} {Foundations of Physics}\ }\textbf {\bibinfo
  {volume} {51}},\ \bibinfo {pages} {30} (\bibinfo {year} {2021})},\ \Eprint
  {https://arxiv.org/abs/2006.15543} {arXiv:2006.15543} \BibitemShut {NoStop}%
\bibitem [{\citenamefont {Drezet}(2022)}]{Drezet}%
  \BibitemOpen
  \bibfield  {author} {\bibinfo {author} {\bibfnamefont {A.}~\bibnamefont
  {Drezet}},\ }\bibfield  {title} {\bibinfo {title} {In defense of relational
  quantum mechanics: A note on `relative facts do not exist. relational quantum
  mechanics is incompatible with quantum mechanics'},\ }\Eprint
  {https://arxiv.org/abs/2209.01237} {arXiv:2209.01237}  (\bibinfo {year}
  {2022})\BibitemShut {NoStop}%
\bibitem [{\citenamefont {Lawrence}\ \emph
  {et~al.}(2022{\natexlab{b}})\citenamefont {Lawrence}, \citenamefont
  {Markiewicz},\ and\ \citenamefont {{\.Z}ukowski}}]{LMZ2}%
  \BibitemOpen
  \bibfield  {author} {\bibinfo {author} {\bibfnamefont {J.}~\bibnamefont
  {Lawrence}}, \bibinfo {author} {\bibfnamefont {M.}~\bibnamefont
  {Markiewicz}},\ and\ \bibinfo {author} {\bibfnamefont {M.}~\bibnamefont
  {{\.Z}ukowski}},\ }\bibfield  {title} {\bibinfo {title} {Relative facts do
  not exist. relational quantum mechanics is incompatible with quantum
  mechanics. response to the critique by aur\'lien drezet},\ }\Eprint
  {https://arxiv.org/abs/2210.09025} {arXiv:2210.09025}  (\bibinfo {year}
  {2022}{\natexlab{b}})\BibitemShut {NoStop}%
\bibitem [{\citenamefont {Greenberger}\ \emph {et~al.}(1990)\citenamefont
  {Greenberger}, \citenamefont {Horne}, \citenamefont {Shimony},\ and\
  \citenamefont {Zeilinger}}]{greenberger1990bell}%
  \BibitemOpen
  \bibfield  {author} {\bibinfo {author} {\bibfnamefont {D.~M.}\ \bibnamefont
  {Greenberger}}, \bibinfo {author} {\bibfnamefont {M.~A.}\ \bibnamefont
  {Horne}}, \bibinfo {author} {\bibfnamefont {A.}~\bibnamefont {Shimony}},\
  and\ \bibinfo {author} {\bibfnamefont {A.}~\bibnamefont {Zeilinger}},\
  }\bibfield  {title} {\bibinfo {title} {Bell's theorem without inequalities},\
  }\href {https://doi.org/10/dvzgd5} {\bibfield  {journal} {\bibinfo  {journal}
  {American Journal of Physics}\ }\textbf {\bibinfo {volume} {58}},\ \bibinfo
  {pages} {1131} (\bibinfo {year} {1990})}\BibitemShut {NoStop}%
\bibitem [{\citenamefont {Everett}(1957)}]{Everett:1957hd}%
  \BibitemOpen
  \bibfield  {author} {\bibinfo {author} {\bibfnamefont {H.}~\bibnamefont
  {Everett}},\ }\bibfield  {title} {\bibinfo {title} {{Relative state
  formulation of quantum mechanics}},\ }\href@noop {} {\bibfield  {journal}
  {\bibinfo  {journal} {Rev. Mod. Phys.}\ }\textbf {\bibinfo {volume} {29}},\
  \bibinfo {pages} {454} (\bibinfo {year} {1957})}\BibitemShut {NoStop}%
\bibitem [{\citenamefont {Bong}\ \emph {et~al.}(2020)\citenamefont {Bong},
  \citenamefont {{Utreras-Alarc{\'o}n}}, \citenamefont {Ghafari}, \citenamefont
  {Liang}, \citenamefont {Tischler}, \citenamefont {Cavalcanti}, \citenamefont
  {Pryde},\ and\ \citenamefont {Wiseman}}]{bong2020strong}%
  \BibitemOpen
  \bibfield  {author} {\bibinfo {author} {\bibfnamefont {K.-W.}\ \bibnamefont
  {Bong}}, \bibinfo {author} {\bibfnamefont {A.}~\bibnamefont
  {{Utreras-Alarc{\'o}n}}}, \bibinfo {author} {\bibfnamefont {F.}~\bibnamefont
  {Ghafari}}, \bibinfo {author} {\bibfnamefont {Y.-C.}\ \bibnamefont {Liang}},
  \bibinfo {author} {\bibfnamefont {N.}~\bibnamefont {Tischler}}, \bibinfo
  {author} {\bibfnamefont {E.~G.}\ \bibnamefont {Cavalcanti}}, \bibinfo
  {author} {\bibfnamefont {G.~J.}\ \bibnamefont {Pryde}},\ and\ \bibinfo
  {author} {\bibfnamefont {H.~M.}\ \bibnamefont {Wiseman}},\ }\bibfield
  {title} {\bibinfo {title} {A strong no-go theorem on the {{Wigner}}'s friend
  paradox},\ }\href {https://doi.org/10/gg85dd} {\bibfield  {journal} {\bibinfo
   {journal} {Nature Physics}\ ,\ \bibinfo {pages} {1}} (\bibinfo {year}
  {2020})}\BibitemShut {NoStop}%
\bibitem [{\citenamefont {Peres}(1978)}]{Peres1978}%
  \BibitemOpen
  \bibfield  {author} {\bibinfo {author} {\bibfnamefont {A.}~\bibnamefont
  {Peres}},\ }\bibfield  {title} {\bibinfo {title} {Unperformed experiments
  have no results},\ }\href {https://doi.org/10.1119/1.11393} {\bibfield
  {journal} {\bibinfo  {journal} {American Journal of Physics}\ }\textbf
  {\bibinfo {volume} {46}},\ \bibinfo {pages} {745} (\bibinfo {year}
  {1978})}\BibitemShut {NoStop}%
\bibitem [{\citenamefont {Adlam}\ and\ \citenamefont
  {Rovelli}(2022)}]{adlam2022information}%
  \BibitemOpen
  \bibfield  {author} {\bibinfo {author} {\bibfnamefont {E.}~\bibnamefont
  {Adlam}}\ and\ \bibinfo {author} {\bibfnamefont {C.}~\bibnamefont
  {Rovelli}},\ }\bibfield  {title} {\bibinfo {title} {Information is
  {{Physical}}: {{Cross-Perspective Links}} in {{Relational Quantum
  Mechanics}}},\ }\Eprint {https://arxiv.org/abs/2203.13342} {arXiv:2203.13342}
   (\bibinfo {year} {2022})\BibitemShut {NoStop}%
\bibitem [{\citenamefont {Brukner}(2018)}]{brukner2018nogo}%
  \BibitemOpen
  \bibfield  {author} {\bibinfo {author} {\bibfnamefont {{\v C}.}~\bibnamefont
  {Brukner}},\ }\bibfield  {title} {\bibinfo {title} {A {{No-Go Theorem}} for
  {{Observer-Independent Facts}}},\ }\href {https://doi.org/10/gdq8td}
  {\bibfield  {journal} {\bibinfo  {journal} {Entropy}\ }\textbf {\bibinfo
  {volume} {20}},\ \bibinfo {pages} {350} (\bibinfo {year} {2018})},\ \Eprint
  {https://arxiv.org/abs/1804.00749} {arXiv:1804.00749} \BibitemShut {NoStop}%
\bibitem [{\citenamefont {Cavalcanti}\ and\ \citenamefont
  {Wiseman}(2021)}]{cavalcanti2021implicationsa}%
  \BibitemOpen
  \bibfield  {author} {\bibinfo {author} {\bibfnamefont {E.~G.}\ \bibnamefont
  {Cavalcanti}}\ and\ \bibinfo {author} {\bibfnamefont {H.~M.}\ \bibnamefont
  {Wiseman}},\ }\bibfield  {title} {\bibinfo {title} {Implications of {{Local
  Friendliness}} violation for quantum causality},\ }\href
  {https://doi.org/10/gm7w6p} {\bibfield  {journal} {\bibinfo  {journal}
  {Entropy}\ }\textbf {\bibinfo {volume} {23}},\ \bibinfo {pages} {925}
  (\bibinfo {year} {2021})},\ \Eprint {https://arxiv.org/abs/2106.04065}
  {arXiv:2106.04065} \BibitemShut {NoStop}%
\bibitem [{\citenamefont {Haddara}\ and\ \citenamefont
  {Cavalcanti}(2022)}]{Haddara2022}%
  \BibitemOpen
  \bibfield  {author} {\bibinfo {author} {\bibfnamefont {M.}~\bibnamefont
  {Haddara}}\ and\ \bibinfo {author} {\bibfnamefont {E.~G.}\ \bibnamefont
  {Cavalcanti}},\ }\bibfield  {title} {\bibinfo {title} {A possibilistic no-go
  theorem on the wigner's friend paradox},\ }\Eprint
  {https://arxiv.org/abs/2205.12223} {arXiv:2205.12223}  (\bibinfo {year}
  {2022})\BibitemShut {NoStop}%
\end{thebibliography}%

\end{document}